\documentclass[reprint,aps,prb,superscriptaddress]{revtex4-1}

\usepackage{graphicx}
\usepackage{latexsym}
\usepackage{siunitx}
\usepackage{amsmath, array, makecell}
\usepackage{cellspace} 
\begin{document}

\title{\LARGE Analysis and mitigation of interface losses in trenched superconducting coplanar waveguide resonators}

\author{G. Calusine}
\thanks{greg.calusine@ll.mit.edu; These authors contributed equally to this work.}
\affiliation{MIT Lincoln Laboratory, 244 Wood Street, Lexington, MA 02421, USA}
\author{A. Melville}
\thanks{greg.calusine@ll.mit.edu; These authors contributed equally to this work.}
\affiliation{MIT Lincoln Laboratory, 244 Wood Street, Lexington, MA 02421, USA}
\author{W. Woods}
\thanks{greg.calusine@ll.mit.edu; These authors contributed equally to this work.}
\affiliation{MIT Lincoln Laboratory, 244 Wood Street, Lexington, MA 02421, USA}
\author{R. Das}
\affiliation{MIT Lincoln Laboratory, 244 Wood Street, Lexington, MA 02421, USA}
\author{C. Stull}
\affiliation{MIT Lincoln Laboratory, 244 Wood Street, Lexington, MA 02421, USA}
\author{V. Bolkhovsky}
\affiliation{MIT Lincoln Laboratory, 244 Wood Street, Lexington, MA 02421, USA}
\author{D. Braje}
\affiliation{MIT Lincoln Laboratory, 244 Wood Street, Lexington, MA 02421, USA}
\author{D. Hover}
\affiliation{MIT Lincoln Laboratory, 244 Wood Street, Lexington, MA 02421, USA}
\author{D. K. Kim}
\affiliation{MIT Lincoln Laboratory, 244 Wood Street, Lexington, MA 02421, USA}
\author{X. Miloshi}
\affiliation{MIT Lincoln Laboratory, 244 Wood Street, Lexington, MA 02421, USA}
\author{D. Rosenberg}
\affiliation{MIT Lincoln Laboratory, 244 Wood Street, Lexington, MA 02421, USA}
\author{A. Sevi}
\affiliation{MIT Lincoln Laboratory, 244 Wood Street, Lexington, MA 02421, USA}
\author{J. L. Yoder}
\affiliation{MIT Lincoln Laboratory, 244 Wood Street, Lexington, MA 02421, USA}
\author{E. Dauler}
\affiliation{MIT Lincoln Laboratory, 244 Wood Street, Lexington, MA 02421, USA}
\author{W. D. Oliver}
\affiliation{MIT Lincoln Laboratory, 244 Wood Street, Lexington, MA 02421, USA}
\affiliation{Research Laboratory of Electronics, Massachusetts Institute of Technology, Cambridge, MA 02139, USA}
\affiliation{Department of Physics, Massachusetts Institute of Technology, Cambridge, MA 02139, USA}

\begin{abstract}
Improving the performance of superconducting qubits and resonators generally results from a combination of materials and fabrication process improvements and design modifications that reduce device sensitivity to residual losses.  One instance of this approach is to use trenching into the device substrate in combination with superconductors and dielectrics with low intrinsic losses to improve quality factors and coherence times.  Here we demonstrate titanium nitride coplanar waveguide resonators with mean quality factors exceeding two million and controlled trenching reaching \SI{2.2}{\micro\meter} into the silicon substrate.  Additionally, we measure sets of resonators with a range of sizes and trench depths and compare these results with finite-element simulations to demonstrate quantitative agreement with a model of interface dielectric loss.  We then apply this analysis to determine the extent to which trenching can improve resonator performance.   
\end{abstract}

\maketitle

Dielectric loss associated with two-level systems (TLS) at materials interfaces is a major contributor limiting coherence times and quality factors in superconducting qubit and resonator devices.\cite{Oliver2013,Martinis2005,Gao2008,Gao2008a,Sage2011}  In order to mitigate these losses, previous work has employed a combination of improving materials, optimizing fabrication, and modifying designs.\cite{Khalil2011,Wang2015,Chu2016,Dial2016,Gambetta2017}  Materials and fabrication efforts have focused primarily on lowering the density of TLS defects in bulk materials\cite{Paik2010}, and reducing the presence of TLS-containing dielectrics\cite{OConnell2008} and chemical residues.\cite{Quintana2014}  Device geometry and design parameter modifications have in turn been used to reduce device sensitivity to material losses by tailoring the structure's electromagnetic field profile.\cite{Wenner2011,Koch2007,Yan2016}  Together these advances have yielded qubit $T_1$ times exceeding \SI{50}{\micro\second}\cite{Jin2015,Yan2016,Rigetti2012} and resonator internal quality factors ($Q_i$) reaching 70 million at single photon-excitation powers.\cite{Reagor2016}\par

Despite these remarkable accomplishments, developing a complete understanding of interfacial TLS loss mechanisms has remained a challenge.  For example, although materials with reduced TLS losses such as titanium nitride (TiN) have been used to realize high $Q_i$ resonators and long $T_1$ qubits,\cite{Chang2013} the results often exhibit poor reproducibility in part due to the metal's sensitivity to ambient oxygen.\cite{Ohya2014}  Additionally, while improvements in $T_1$ and $Q_i$ have been demonstrated through the use of substrate trenching to reduce interface participation,\cite{Vissers2012a,Bruno2015,Gambetta2017} the depth dependence and the degree to which deeper trenching improves device performance remains unclear.  Finite-element electromagnetic modeling of dielectric losses can be used to study these effects, but it must be paired with highly controllable and reproducible fabrication processes to make quantitative comparisons between experiment and simulations. \par

In this work, we present TiN coplanar waveguide (CPW) resonators with quality factors exceeding two million fabricated using a process capable of controlled trenching in the silicon substrate.  To analyze losses in these devices, we perform finite-element electromagnetic simulations of a range of resonator geometries in order to analyze interfacial and substrate dielectric losses.  We then demonstrate quantitative agreement between measured CPW resonator $Q_i$'s and a model of interface losses.  Furthermore, we use this tool to predict the marginal benefits of deep trenching for reducing losses in superconducting CPW resonators.  The agreement supports the accuracy of interface participation ratio-based modeling of device losses and indicates future pathways for reducing loss in superconducting devices. \par

\begin{figure*}
\includegraphics[scale=.64]{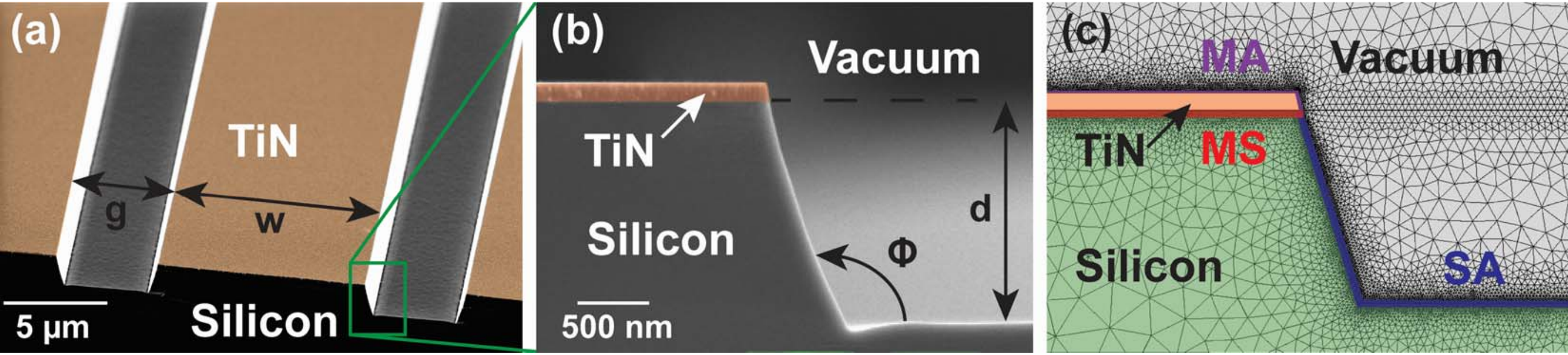}
\caption{\label{fig1} (a) Scanning electron microscope (SEM) image of a representative TiN (false-colored orange) resonator with width (\textit{w}) and gap (\textit{g}) to ground plane.  (b) Cross-sectional SEM image of the same TiN resonator with trench depth (\textit{d}) and sidewall angle ($\mathit{\Phi}$).  (c) 2D finite-element mesh used to calculate participation ratios with the dielectric regions false-colored as follows: the metal-to-substrate interface (MS, red), the substrate-to-air/vacuum interface (SA, blue), the metal-to-air/vacuum interface (MA, purple), and the bulk silicon substrate (Si, green).}
\end{figure*}

We study superconducting CPW quarter-wave resonators with a center trace width \textit{w} ranging from \SI{3}{\micro\meter} to \SI{22}{\micro\meter} and gap \textit{g} to ground ranging from \SI{1.5}{\micro\meter} to \SI{11}{\micro\meter} [see Fig. 1(a)].  The devices were fabricated using a subtractive etch process on high resistivity 200 mm (001) silicon substrates ($\geq$ 3500 $\Omega$-cm).  Prior to metal deposition, the substrates were prepared using an RCA clean in conjunction with megasonication. Without additional oxide removal steps or buffer layers, we reactively sputtered 150 nm of TiN using a titanium target in the presence of argon and nitrogen gas. We patterned the resonators using optical lithography and then etched the metal and underlying substrate using a combination of BCl$_3$ and Cl$_2$ gases. The total etch time was adjusted to control the trench depth (\textit{d}).  We then used an \textit{in situ} oxygen plasma ash followed by an \textit{ex situ} hydroxylamine-based wet strip to remove the remaining photoresist.  Figure 1(b) shows a representative CPW resonator cross section.  With the sole exception of the variable etch time, we use a nominally identical fabrication process for all samples and therefore attribute differences in $Q_i$ to the trench depth and not to changes in the interfacial loss tangents.  Further details of the chip design and fabrication process are provided in the supplementary material.\cite{Supplemental}\par

TLS losses in superconducting CPW resonators can be understood by applying an interface participation ratio model similar to those used in References \onlinecite{Wenner2011,Wang2015,Dial2016,Gambetta2017}.   In this model, the resonator dielectric losses are a linear combination of the loss tangents ($\tan{\delta_i}$) associated with energy absorbing TLS's in each region $i$, weighted by the fraction of the total electric field energy stored in that region, the participation ratio $p_i$: 
\begin{equation}
\frac{1}{Q_{TLS}}=\sum\limits_{i}{p_i\tan{\delta_i}}
\end{equation}
Because each lossy region contains an unknown combination of interface dielectrics and fabrication residues, in our analysis, we assign a unique $\tan{\delta_i}$ to each interface that is exposed to a distinct fabrication process.  The participation ratios of the dielectric regions in our devices were calculated using two-dimensional (2D) COMSOL electrostatic simulations\cite{COMSOL}.  We partition the device into the following lossy dielectric regions: the metal-to-silicon (MS, red), substrate-to-air/vacuum (SA, blue), and metal-to-air/vacuum (MA, purple) interfaces, and the bulk silicon substrate (Si, green), as depicted by the false coloring in Fig. 1(c). To reduce the computational complexity, the interface participation ratio calculations were performed using 10 nm thick defect layers of a fixed dielectric constant $\epsilon$ = 10, despite general uncertainty in the actual interface properties.  This results in ambiguity in the resulting values for $\tan{\delta_i}$.  However, due to the manner in which participation ratios scale with thickness and dielectric constant in the limit of a thin layer\cite{Wenner2011}, we can parameterize $1/Q_{TLS}$ in Eq. (1) using scaled participation ratios $P_i$ and `loss factors' $x_i$ that are independent of these quantities and are defined by $\sum\limits_{i}{p_i\tan{\delta_i}}= \sum\limits_{i}{P_ix_i}$.  For details, see the supplementary materials\cite{Supplemental}.  For all resonator geometries, the trench sidewall angle $\mathit{\Phi}$ and depth $\mathit{d}$ were determined using cross-sectional scanning electron microscopy in order to accurately model the device electric field distribution.  All devices exhibit angled sidewalls with $\mathit{\Phi}$ ranging from 93-109$^{\circ}$ depending on the etch time and feature size. \par

 \begin{figure*}
\includegraphics[scale=.64]{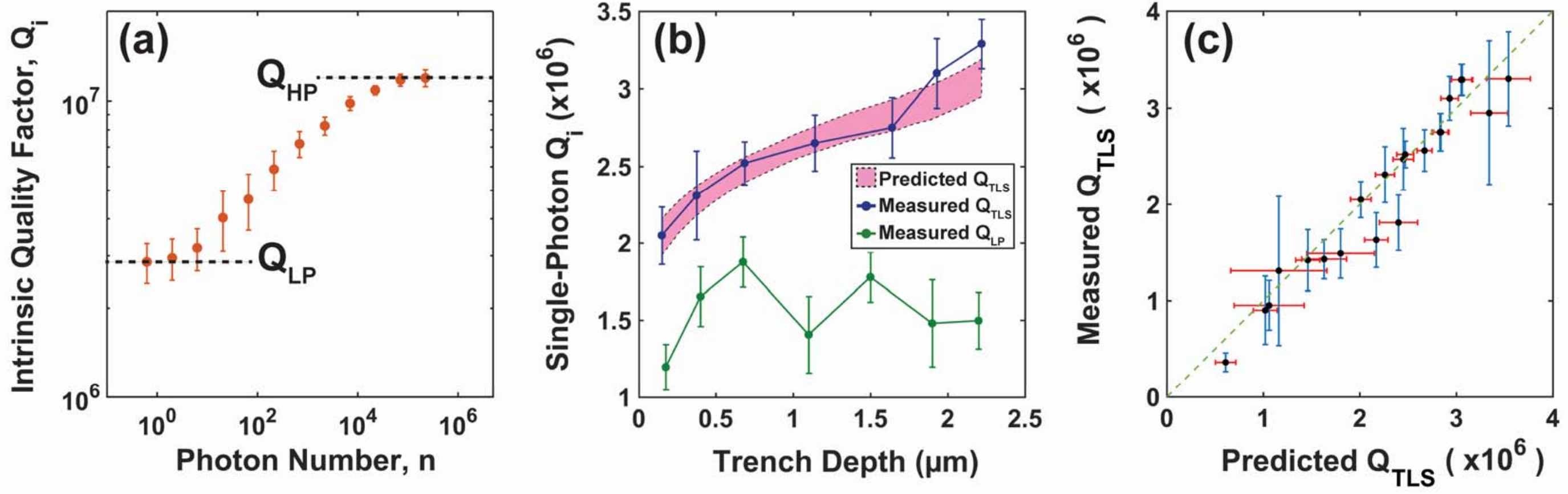}
\caption{\label{fig3} (a) Representative intrinsic quality factor ($Q_i$) as a function of photon number for a resonator with (\textit{w}, \textit{g}, \textit{d}) = (\SI{16}{\micro\meter}, \SI{8}{\micro\meter}, \SI{0.68}{\micro\meter}). The low power and high power limits are indicated with dashed lines. (b) $Q_i$ as a function of trench depth with the same CPW geometry as in Fig. 2(a). Each data point represents the mean $Q_i$ obtained from 10-15 nominally identical resonators. The green data points represent the low-power quality factor ($Q_{LP}$), and the blue data points represent TLS-limited quality factors ($Q_{TLS}$). The pink shaded region represents the 95\% confidence interval of the predicted $Q_{TLS}$. (c) Predicted $Q_i$ (red error bars) compared to measured $Q_{TLS}$ (blue error bars).  The dashed line corresponds the ideal case where the two values are equal.  All error bars represent 95\% confidence intervals.}
\end{figure*}

To compare our interface loss simulations to our fabricated device performance, we characterized a series of resonators with a range of geometries by measuring the resonator chip transmission spectrum at 25 mK in a dilution refrigerator.  Background ambient magnetic fields were reduced by mounting the package in a superconducting aluminum enclosure surrounded by a high magnetic permeability shield (Cryoperm).  In each cooldown, twelve packages each containing a single chip comprising five resonators were measured using two separate measurement chains that incorporated a pair of $1\times6$ microwave switches operating at the base temperature stage of the dilution refrigerator.  Each measurement chain included a series of microwave attenuators, filters, and isolators to reduce the samples' exposure to thermal radiation from hotter temperature stages.  A broadband traveling wave parametric amplifier,\cite{Macklin2015} low-noise high-electron-mobility-transistor amplifier, and room temperature microwave amplifier were used to amplify the transmitted signal before measurement using a vector network analyzer.  Each resonator was measured over a range of internal circulating powers from the single-photon limit up to approximately $10^6$ photons using a non-linear frequency spacing to minimize data acquisition times.  Resonator parameters were extracted using the fitting methods presented in Ref. \onlinecite{gaothesis}.  Each device was measured repeatedly in the single-photon limit for approximately five hours, and the results were averaged in order to account for time-dependent $Q_i$ fluctuations.  Similarly, multiple copies of the same device were measured to account for device-to-device variations and to establish error bars for each sample set. \par

Figure 2(a) shows an example of the dependence of $Q_i$ on the number of photons circulating in the resonator.  The trend shows the typical saturation behavior of internal losses associated with TLS's. At low internal photon numbers(circulating power), the $Q_i$, which we label $Q_{LP}$, is dominated by absorption due to unsaturated TLS's.\cite{Gao2008}  At higher photon numbers, this loss mechanism saturates and $Q_i$ increases until it reaches another limiting value $Q_{HP}$.  At this power, the losses cease to be dominated by TLS's and are instead dominated by an unknown combination of other mechanisms such as vortices\cite{Song_2009,Nsanzineza2014}, radiation/packaging loss\cite{Sage2011}, and/or non-equilibrium quasiparticles.\cite{Visser2011}  The $Q_{LP}$ and $Q_{HP}$ shown in Fig. 2(a) are typical of our highest mean $Q_i$ fabrication process with mean $Q_i$ of $2.2\times10^6$ for a sample set of 15 resonators with ($w$, $g$, $d$) = (\SI{16}{\micro\meter}, \SI{8}{\micro\meter}, \SI{0.68}{\micro\meter}).  To study the effects of trenching down to \SI{2.2}{\micro\meter} depth, we instead use a similar process with a thicker  photoresist mask (\SI{4}{\micro\meter} vs. \SI{1.1}{\micro\meter}) and a higher temperature post-etch ash.  This leads to an approximate and reproducible 15\% reduction in mean $Q_i$ for a comparable set of devices with this resonator geometry and trench depth.  To assess the reproducibility of this fabrication process, for the shallowest trenching shown here (150 nm), we have measured approximately 100 nominally identical resonators and observed that greater than 87\% show $Q_i$'s higher than $1\times10^6$ (mean of $1.6\times10^6$). \cite{Supplemental} \par
Although TLS's are generally the dominant source of loss in superconducting CPW resonators at low temperature and circulating power, the losses that persist when TLS's are saturated can still reduce total $Q_i$ and contribute to device-to-device variation.  All of the resonators we characterized exhibited TLS-saturation behavior similar to the data shown in Fig. 2(a), yet we observed significant variation in $Q_{HP}$.  As a result, the differences we observed in $Q_{LP}$ were sometimes dominated by $Q_{HP}$ variations rather than altered interface participation.   This resulted in behavior such as shown in Fig. 2(b) (green points and lines) where no discernible trend in $Q_{LP}$ vs. trench depth is observed.  However, since we can separately measure $Q_{HP}$ by saturating the losses associated with $Q_{TLS}$, we can subtract the contributions from other loss mechanisms to determine $Q_{TLS}$ from $Q_{LP}$:
 \begin{equation}
 \frac{1}{Q_{TLS}}=\frac{1}{Q_{LP}}-\frac{1}{Q_{HP}}
 \end{equation}
The blue points and lines in Fig. 2(b) show the $Q_{TLS}$ values determined from the $Q_{LP}$ values (green points and lines) vs. trench depth when this correction is performed independently for each device in the dataset.  This data set exhibits the expected monotonic improvement in $Q_i$ as interface participation ratios decrease with increasing trench depths.  The error bars represent 95\% confidence intervals resulting from measuring 10-15 nominally identical devices for each depth.\par

\begin{figure*}
\includegraphics[scale=.64]{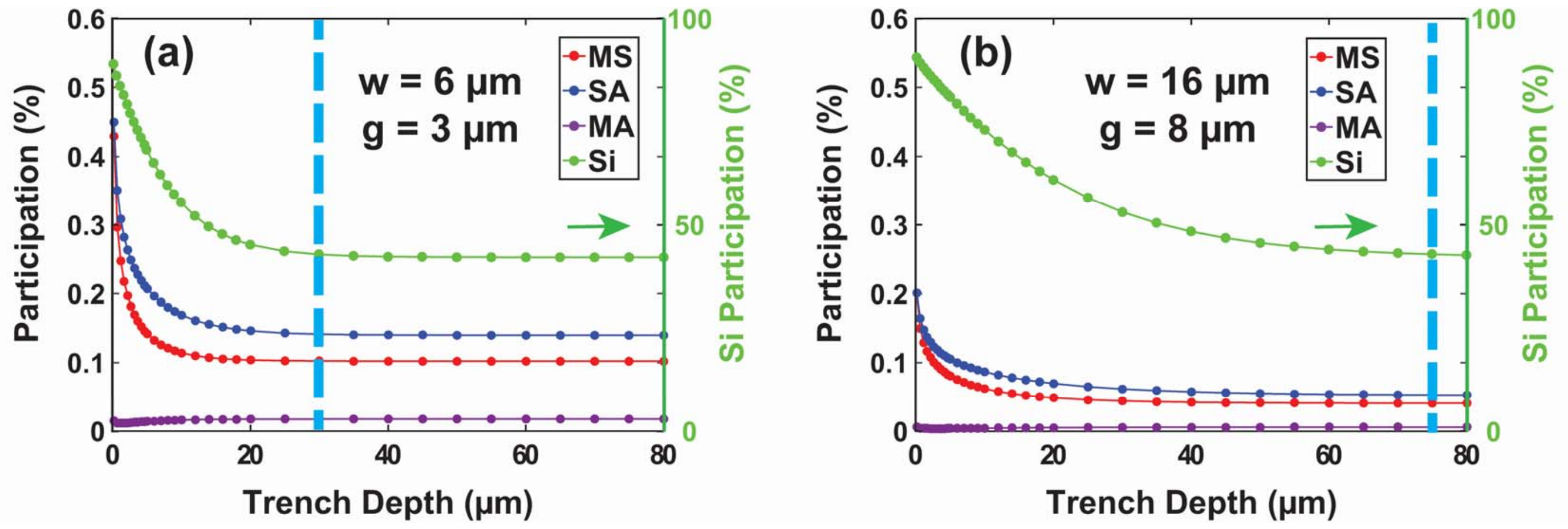}
\caption{\label{fig2} Participation ratios as a function of trench depth for two representative resonator geometries (a) (\textit{w}, \textit{g}) = (\SI{6}{\micro\meter}, \SI{3}{\micro\meter}), and (b) (\textit{w}, \textit{g}) = (\SI{16}{\micro\meter}, \SI{8}{\micro\meter}). MS (red), SA (blue), and MA (purple) participation ratios are plotted on the left axis, and the Si participation ratio (green) is plotted on the right axis.}
\end{figure*}

In order to develop a quantitative model of interface losses in our devices, we additionally characterized a series of resonator geometries ranging from (\textit{w}, \textit{g}) = (\SI{3}{\micro\meter}, \SI{1.5}{\micro\meter}) to (\textit{w}, \textit{g}) = (\SI{22}{\micro\meter}, \SI{11}{\micro\meter}) for trench depths between \SI{0.15}{\micro\meter} and \SI{2.2}{\micro\meter}.  Each geometry in this dataset provides a linear equation of the form of Eq. (1) relating device $Q_{TLS}$ to loss factors associated with each dielectric region.  While a single relationship is insufficient to determine each region's losses, multiple geometries with varying combinations of participation ratios form a set of linear equations that can in principle be used to determine each individual loss factor.  However, in general, this matrix of participation ratios is very nearly singular for a wide range of planar geometries.   This collinearity is readily apparent in the approximate proportionality of the MS and SA interface participation ratios at all depths shown in Fig. 3.  As a result, errors associated with the input $Q_{TLS}$ values and modeling inaccuracy prevent the determination of a unique solution to the system of equations.   Nevertheless, we can perform a Monte Carlo analysis of the constrained least square optimization solution using our measured $Q_{TLS}$ values and error bars in order to determine a corresponding distribution of loss factors for the dielectric regions. For comparisons to previously reported loss tangents, see the supplementary materials\cite{Supplemental}.  A comparison between the measured $Q_{TLS}$ and the $Q_{TLS}$ predicted by this model is shown in Fig. 2(c) with the corresponding error bars for the 19 device geometries that we measured. The dashed green line represents the values where the measured and predicted $Q_{TLS}$ correspond exactly.  This model can also be used to determine predictive bounds for resonator $Q_{TLS}$ for devices with other geometries and trench depths.  The region of 95\% confidence in this prediction is shown in Fig. 2(b) for (\textit{w}, \textit{g}) = (\SI{16}{\micro\meter}, \SI{8}{\micro\meter}) devices (pink shaded region) over the range of trench depths we studied.  The predicted $Q_{TLS}$ agree well with the measured values, indicating that the interface losses are likely uniform between resonators with different trench depths.  \par

To determine the extent to which $Q_{TLS}$ can be improved with increasing trench depth, we simulate the interface participation ratios for depths comparable to those achievable through deep silicon etching\cite{Bruno2015} for multiple geometries and assuming perpendicular sidewall angles ($\mathit{\Phi}$ = 90$^{\circ}$). Figs. 3(a) and 3(b) show MS (red), SA (blue), MA (purple), and Si (green) participation ratios for two representative coplanar resonator geometries (\textit{w}, \textit{g}) = (\SI{6}{\micro\meter}, \SI{3}{\micro\meter}) and (\SI{16}{\micro\meter}, \SI{8}{\micro\meter}) as a function of trench depth from \textit{d} = \SI{0.15}{\micro\meter} to \textit{d} = \SI{80}{\micro\meter}. The interface participation decreases with trench depth, and it asymptotes beyond a depth that is dependent on the CPW gap.  The blue dashed line indicates the depth at which the total bulk and interface participation reaches within 1\% of the asymptotic value. In general, we observe that trenching beyond a depth of approximately \textit{d} = 10\textit{g} ceases to further reduce the participation ratios in the device interfaces or the silicon substrate.  This asymptotic behavior can be contrasted with the logarithmic dependence at 1-\SI{10}{\micro\meter} depths simulated in Ref. \onlinecite{Gambetta2017}. \par

In summary, we have demonstrated trenched TiN resonators with a mean $Q_i$ of 2.2 million.  Characterization of sets of devices with a range of CPW dimensions and trench depths has enabled us to produce a model of dielectric losses that quantitatively agrees with our measured $Q_i$'s and can be used to predict device performance within the bounds set by the model uncertainty.  Furthermore, we have used this form of participation ratio-based device modeling to predict the extent to which deep trenching can improve dielectric losses in superconducting CPW resonators.  Altogether these results indicate that trenching significantly reduces aggregate interface dielectric losses in superconducting CPW resonators and that significant further improvements in total $Q_i$ are possible by mitigating loss contributions from non-TLS related sources.  Additionally, it may be possible to combine the analysis method we use to model dielectric losses in our system with more drastic geometry changes in order to more accurately determine interface losses as a tool for process qualification and device improvement.  Both approaches would provide essential information for reducing dielectric losses in superconducting quantum devices.  \par

We gratefully acknowledge M. Augeri, J. Birenbaum, P. Baldo, M. Cook, G. Fitch, E. Golden, V. Iaia, K. Magoon, P. Murphy, B. Osadchy, A. Sears, R. Slattery, C. Thoummaraj, and D. Volfson at MIT Lincoln Laboratory for technical assistance. This material is based upon work supported by the Department of Defense under Air Force Contract No. FA8721-05-C-0002 and/or FA8702-15-D-0001. Any opinions, findings, conclusions or recommendations expressed in this material are those of the authors and do not necessarily reflect the views of the Department of Defense.

\pagebreak
\widetext
\begin{center}
\textbf{\LARGE Supplementary Materials for ``Analysis and mitigation of interface losses in trenched superconducting coplanar waveguide resonators"\\}
\end{center}
\twocolumngrid

\setcounter{equation}{0}
\setcounter{figure}{0}
\setcounter{table}{0}
\setcounter{page}{1}
\renewcommand{\thefigure}{S\arabic{figure}}
\renewcommand{\theequation}{S\arabic{equation}}
\renewcommand{\bibnumfmt}[1]{[S#1]}
\renewcommand{\citenumfont}[1]{S#1}

\section{Fabrication details}

All devices reported in the main text were fabricated on high resistivity 200 mm (001) Si wafers (\textgreater3500 $\Omega$-cm). Prior to thin film growth,  the bare wafers were first cleaned using a combination of megasonication and RCA clean process.  The wafers were then coated with 150 nm of TiN in a DC reactive magnetron sputtering system with a background pressure in the low $10^{-8}$ torr range.  The deposited metal layer had film stress ($\sigma$ = 0 $\pm$ 150 MPa), similar to other reports of high $Q_i$ TiN resonators. \cite{sOhya2014}  These conditions were used for all devices measured in this work.  This process yielded a typical resistivity variation of approximately 35\% (8.5 to 11.5 ohms/$\Box$) across the entire 200 mm wafer ($<$ 6\% over the center 75mm).  At low temperature, this material exhibited a kinetic inductance of approximately 2.15 picohenries/$\Box$, which was accounted for in device designs in order to achieve the desired device frequencies. \par

\begin{figure} 
\includegraphics[scale=.32]{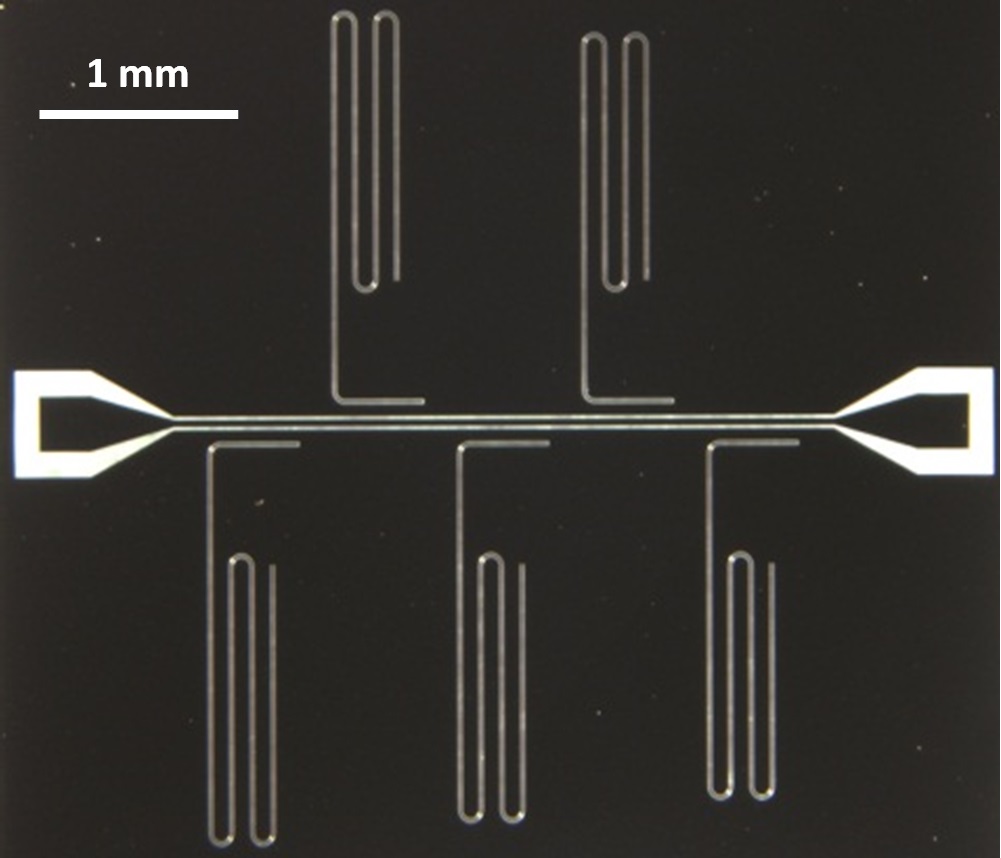}
\caption{Optical image of a typical resonator chip layout used in this work.}
\label{FigS1}
\end{figure}

After deposition, the wafers were patterned using a subtractive etch process and i-line photolithography using a wide field stepper.  We chose to use a thick photoresist instead of a hard mask to maintain a process flow similar to our highest $Q_i$ fabrication process while also allowing for trenching up to a maximum depth of 2.2 $\mu$m. After developing the photoresist, the wafers were etched using BCl$_3$ and Cl$_2$ plasma.  The etch time was varied in order to control the trench depth. Without breaking vacuum, the remaining photoresist was partially stripped using an oxygen plasma. After removing the wafers from the tool, an additional wet chemical strip was used to complete the photoresist removal.  The wafers were then coated in photoresist for dicing into 5 mm by 5 mm chips.  A typical resonator chip is shown in Fig. \ref{FigS1}. \par

The device chips consisted of superconducting coplanar waveguide (CPW) quarter-wave resonators with a center trace width \textit{w} ranging from \SI{3}{\micro\meter} to \SI{22}{\micro\meter} and gap \textit{g} to ground ranging from \SI{1.5}{\micro\meter} to \SI{11}{\micro\meter} [see Fig. 1(a) of the main text].  One end of the resonator is shorted to ground and the other end is isolated with a gap \textit{g} to ground.  Five such resonators are capacitively coupled to a 50 $\Omega$ feedline on each chip.  The resonator lengths are varied to frequency multiplex the resonances in the range of 5-6 GHz with a spacing of approximately 200 MHz.  The resonators are coupled to the central feedline by a $\sim$ \SI{300}{\micro\meter} section of the CPW running parallel to the feedline gap.  The resonator coupling quality factors ($Q_c$) are designed to be comparable to the device $Q_i$ to reduce fitting errors.  Each chip is connected using wirebonds to a gold-plated copper package that contains a microwave feedthrough and an interposer that routes the excitation signal to the chip.  The ground planes on each chip are perforated to trap vortices that might arise due to stray magnetic fields,\cite{sBothner2011,sChiaro2016} and the ground plane is connected to the package ground through many parallel wirebonds.  \par

\begin{figure*} 
\includegraphics[scale=.2]{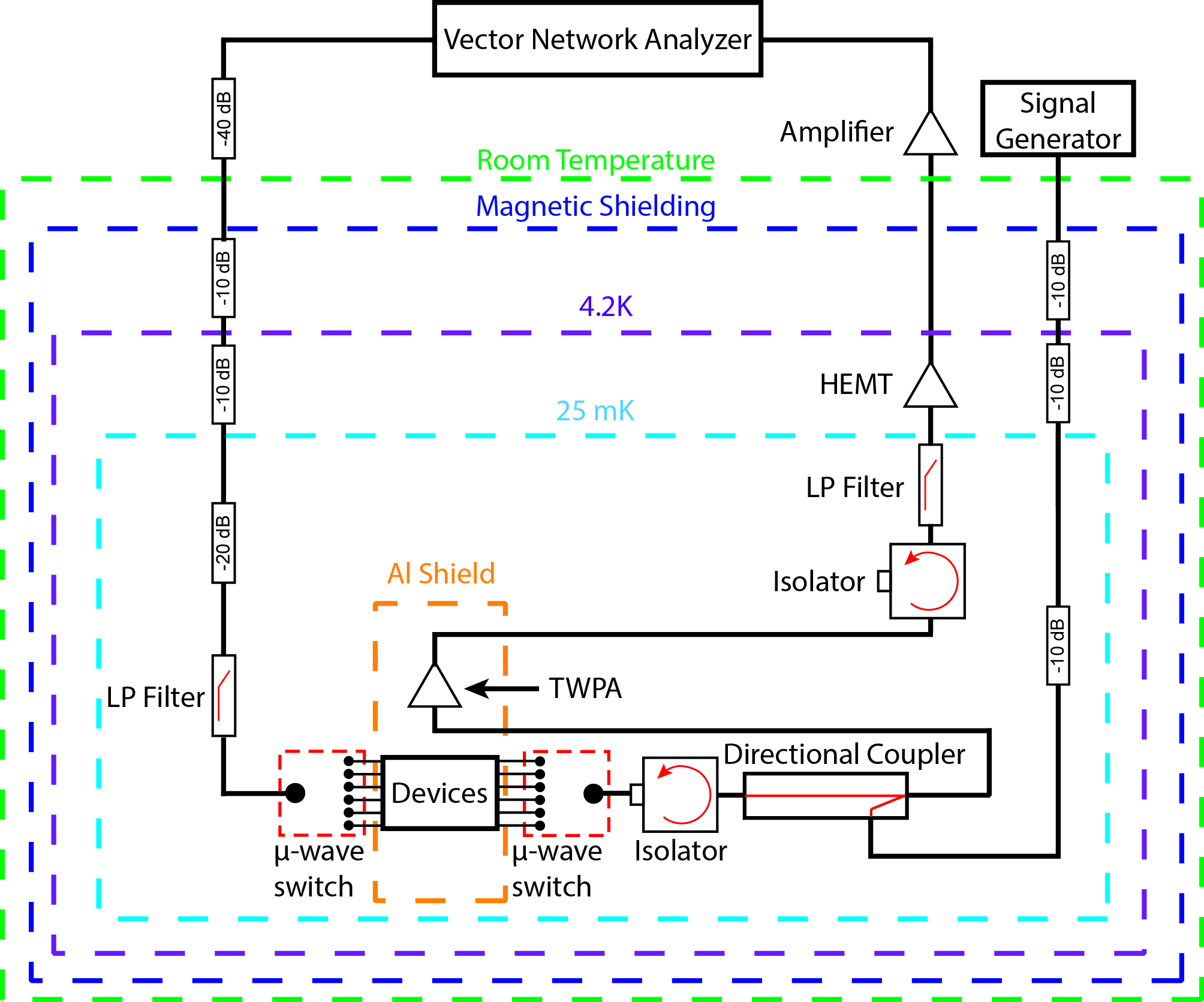}
\caption{Simplified schematic of the resonator measurement setup.}
\label{FigS2}
\end{figure*}

\section{Measurement Hardware}

All measurements in this work were performed in one of two liquid helium dilution refrigerators with a base temperature of 20 mK outfitted with microwave coaxial cabling from room temperature to the base temperature stage. The inner walls of the liquid helium storage dewar were lined with three nested layers of high permeability magnetic shielding in order to expel background magnetic fields.  To further reduce background fields, the sample space was enclosed in a superconducting aluminum shield and only non-magnetic metals and microwave components were used within this enclosure.  This aluminum enclosure was lined with infrared absorbing material to absorb any stray light that could generate non-equilibrium quasiparticles in the devices under test. Twelve device packages each containing a chip with five resonators were cooled down during each measurement cycle.  The base temperature stage of the dilution refrigerator was stabilized at 25 mK throughout the entire measurement process. \par

Figure \ref{FigS2} shows a diagram of the measurement setup used to characterize the devices presented in this work.  A vector network analyzer at room temperature measured the microwave transmission signal through the dilution refrigerator coaxial measurement chain.  The input signal was attenuated by 40 dB at room temperature before undergoing another 40 dB of cryogenic attenuation at various stages of the dilution refrigerator.  The low-temperature cryogenic attenuators and wideband (12-50 GHz) low pass filters were included to prevent thermal radiation from appreciably heating the device-under-test.  A pair of 1 $\times$ 6 microwave switches were used to measure multiple chips at low temperature on each measurement chain during each dilution refrigerator cooldown.  A cryogenic isolator with $\geq$ 36 dB of isolation was placed immediately after the devices to prevent signals from being reflected back towards the devices due to impedance mismatches in the amplification chain.  A directional coupler combined the measurement signal with an off-resonant, continuous wave pump at the input of a Josephson junction traveling wave parametric amplifier in order to achieve approximately 20 dB of nearly quantum-limited amplification.  This amplified signal was then passed to a high-electron mobility transistor amplifier for further amplification.  Another cryogenic isolator with $\geq$ 18 dB of isolation and a wideband low-pass filter were placed between the amplifiers in order to prevent thermal radiation and any reflected signals from reaching the base temperature stage.  Once the signal reached room temperature, it was further amplified prior to measurement using the vector network analyzer.  Each dilution refrigerator contained two separate, identical measurement chains that allowed for 60 total resonators to be measured during each cooldown.  

\section{Participation Ratio Calculations}

The participation of a single dielectric region, $p_i$, is defined in Eq. \ref{eq1}:

\begin{equation} \label{eq1}
p_i = \frac{U_i}{U_{tot}} = \frac{\int_i{\frac{\epsilon_i | E^2 |}{2} }}{\int_V{\frac{\epsilon_i | E^2 |}{2 }}}
\end{equation}

where $U_i$ is the electric field energy stored in region \textit{i}, $U_{tot}$ is the total electric field energy stored in all regions \textit{i}, $E$ is the local electric field, and $\epsilon_i$ is the dielectric constant of the region \textit{i}.  The volume integrals in the numerator and denominator occur over region \textit{i} and the entire volume of interest, respectively.  If the actual dielectric layer participation ratios $p_i$ were known exactly, the TLS-limited Q value, $Q_{TLS}$, for a resonator would be given by Eq. \ref{eq2}: 

\begin{equation} \label{eq2}
\frac{1}{Q_{TLS}}=\sum_i{p_i\tan{\delta_i}}
\end{equation}

where $\tan{\delta_i}$ is the loss tangent of region \textit{i}.  Accurate calculation of participation ratios requires precise knowledge of the permittivity and thickness of each dielectric interface region, and these values are generally unknown. As a result, the loss tangents cannot be derived directly from measurements.\par
Many previous studies use an assumed value for one or both interface parameters when calculating participation ratios. In contrast, by assuming that the defect layers are very thin, and accordingly that the participation ratios scale predictably with thickness and dielectric value \cite{sWenner2011}, we instead introduce `loss factors' $x_i$ as surrogates for loss tangents, as defined in Eq. \ref{eq3} and \ref{eq4}:

\begin{equation} \label{eq3}
x_{i,\parallel} = \frac{\frac{t_i}{t_{nom,i}}}{\frac{\epsilon_{nom,i}}{\epsilon_i}}\tan{\delta_i}
\end{equation}

\begin{equation} \label{eq4}
x_{i,\bot} = \frac{\frac{t_i}{t_{nom,i}}}{\frac{\epsilon_i}{\epsilon_{nom,i}}}\tan{\delta_i}
\end{equation}

\begin{figure*} 
\includegraphics[scale=1.05]{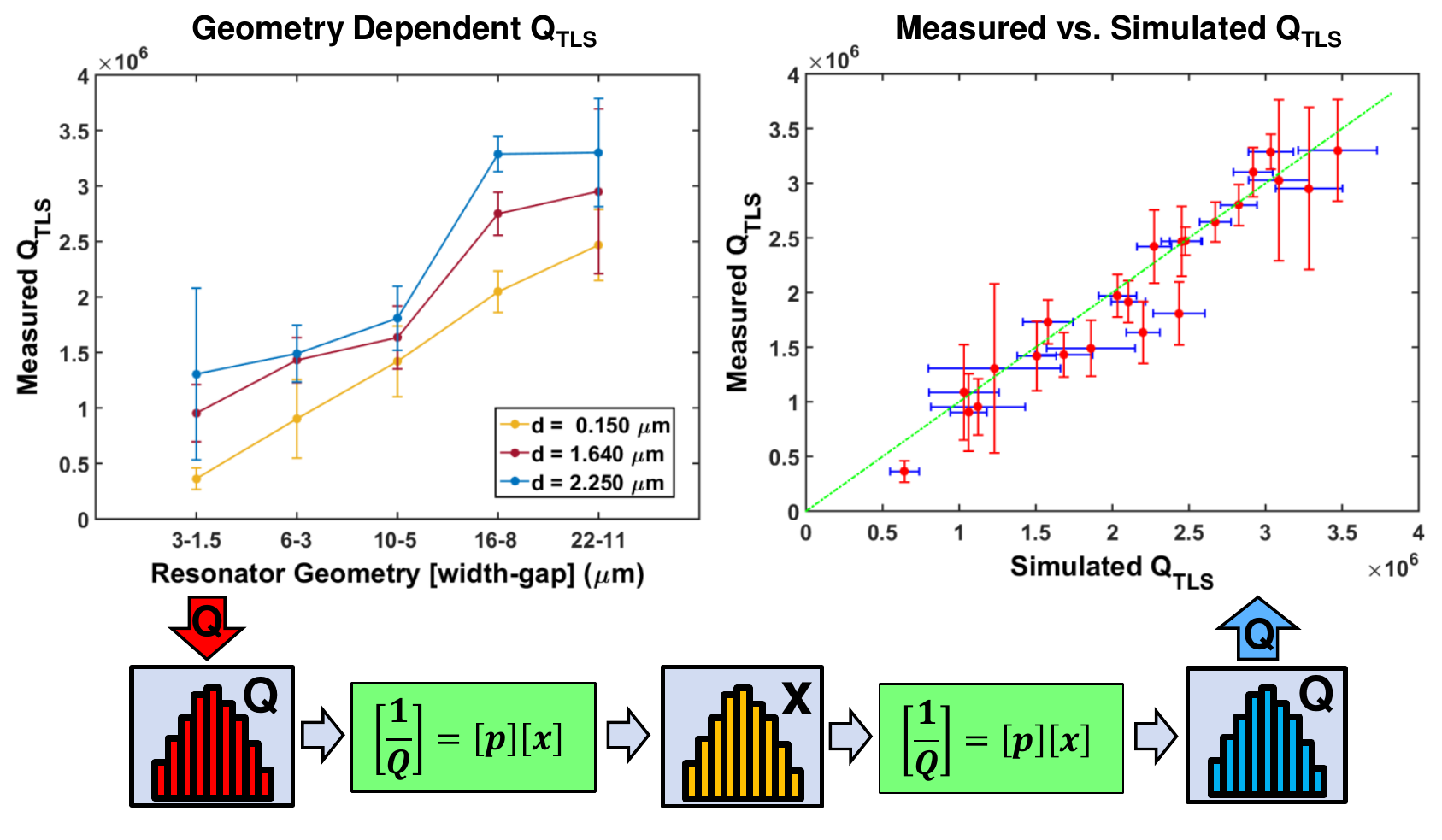}
\caption{Analysis flow showing Monte Carlo extractions of loss-factor vectors, [x] (yellow histogram), from the statistical sampling of measured $Q_{TLS}$ values (red histogram), and converting these extracted loss factor vector distributions to predicted $Q_{TLS}$ values (blue histogram).}
\label{FigS3}
\end{figure*}

Equation \ref{eq3} defines loss factors where the electric field is parallel to the dielectric region, and Eq. \ref{eq4} defines loss factors where the electric field is orthogonal to the dielectric region. In both cases, the loss factors are dimensionless and account for the loss tangents and scaling of the actual defect layer thicknesses and permittivities, $t_i$ and $\epsilon_i$, relative to those used in the participation ratio simulations, $t_{nom,i}$ and $\epsilon_{nom,i}$. The $Q_{TLS}$ of the resonator can then be expressed as a function of simulated participation ratios $P_i$ and loss factors as shown in Eq. \ref{eq5}:

\begin{equation} \label{eq5}
\frac{1}{Q_{TLS}}=\sum_i{P_ix_i}
\end{equation}

Simulated dielectric layer participation ratios for all trenched CPW resonator geometries were obtained using 2D electrostatic simulations in COMSOL.  The center trace width $w$, gap-to-ground $g$, trench depth $d$, and sidewall angle $\mathit{\Phi}$ of all simulated CPW resonators were determined from cross-sectional scanning electron microscope images of fabricated devices.  When simulating geometries at trench depths that were different than those we characterized, the trench sidewall angles were interpolated from surrounding values.  All simulations were performed using a 10 nm thick dielectric layers with  $\epsilon$=10$\epsilon_0$.  The interface dielectrics layers of the CPW structures were partitioned into four regions: metal-to-substrate (MS), substrate-to-air/vacuum (SA), metal-to-air/vacuum (MA), and the silicon substrate (Si) as depicted in Fig. 1(c) of the main text.  The following assumptions were applied for analyzing loss contributions from different surfaces:

\begin{equation} \label{eq7}
P_{MS}\approx P_{MS,\bot} \gg P_{MS,\parallel}
\end{equation}
\begin{equation} \label{eq8}
P_{MA}\approx P_{MA,\bot} \gg P_{MA,\parallel}
\end{equation}
\begin{equation} \label{eq9}
P_{SA}\approx P_{SA,\parallel} \gg P_{SA,\bot}
\end{equation}

Equations \ref{eq7} and \ref{eq8} result from the boundary conditions imposed by assuming perfectly superconducting metals.  Equations \ref{eq9} was derived empirically from our finite-element simulations and was assumed in order to simplify the resulting analysis.  \par

The relations in Eq. \ref{eq7}-\ref{eq9} result in the following approximations for the loss factors:
\begin{equation} \label{eq10}
x_{MS}\approx x_{MS,\bot}
\end{equation}
\begin{equation} \label{eq11}
x_{MA}\approx x_{MA,\bot}
\end{equation}
\begin{equation} \label{eq12}
x_{SA}\approx x_{SA,\parallel}
\end{equation}
\begin{equation} \label{eq13}
x_{Si}\approx \tan{\delta_{Si}}
\end{equation}
$Q_{TLS}$ can then be written as a sum of individual defect layer components as shown in Eq. \ref{eq14}.

\begin{equation} \label{eq14}
\frac{1}{Q_{TLS}}=\frac{1}{Q_{MS}}+\frac{1}{Q_{SA}}+\frac{1}{Q_{MA}}+\frac{1}{Q_{Si}}
\end{equation}

Using Eq. \ref{eq7} through Eq. \ref{eq13}, Eq. \ref{eq14} can be rewritten as shown in Eq. \ref{eq15}:

\begin{equation} \label{eq15}
\frac{1}{Q_{TLS}}=P_{MS}x_{MS}+P_{SA}x_{SA}+P_{MA}x_{MA}+P_{Si}x_{Si}
\end{equation}

The loss factors can be found directly by solving a system of equations of the form of Eq. \ref{eq15} for a series of resonators with simulated dielectric layer participation ratios and the measured $Q_{TLS}$.  This system of equations can be represented in matrix form as shown in Eq. \ref{eq6}:

\begin{equation} \label{eq6}
\Big[\frac{1}{Q_{TLS}}\Big]=[P][x]
\end{equation}

 where [$1/Q_{TLS}$] is a column vector with the number of rows equal to the number of distinct resonator geometries, [\textit{P}] is the participation matrix with the number of rows equal to the number of distinct resonator geometries and the number of columns equal to the number of relevant dielectric regions, and [\textit{x}] is a column vector with the number of rows equal to the number of relevant dielectric regions.\par

We determine the loss factors for our trenched CPW resonator fabrication process by solving Eq. \ref{eq6} for a range of resonator dimensions and trench depths.  The mean resonator $Q_{TLS}$ values are combined with the simulated participation ratios in order to determine a least-squares solution for the loss factor vector [\textit{x}].  In order to determine the uncertainty associated with the output values, we perform a Monte Carlo analysis of the range of output loss factors that result from the mean values and uncertainty of our measured $Q_{TLS}$ values.  Fig. \ref{FigS3} shows a flow diagram depicting how Monte Carlo extractions of loss factor vectors are performed starting from sampling of statistically possible $Q_{TLS}$ and how the output set of extracted loss factor vectors is converted to predicted Q values. The results, presented in Fig. 2(c) of the main text for 10,000 Monte Carlo iterations, show good agreement to the fit of the measured data to the participation model (Eq. \ref{eq15}). The error bars represent the 95\% confidence intervals.\par
The mean best-fit loss factors found when performing the Monte Carlo analysis of the least-squares solutions used to produce the data shown in Fig. 2(c) in the main text are given in Eq. \ref{eq16}: 
\begin{equation} \label{eq16}
\bigg[ \begin{smallmatrix} x_{MS} \\x_{SA} \\x_{MA}  \\x_{Si}\end{smallmatrix} \bigg]_{Mean} = \Bigg[ \begin{smallmatrix} 1.0\times10^{-4} \\5.7\times10^{-5} \\7.8\times10^{-4} \\1.2\times10^{-7}\end{smallmatrix} \Bigg]
\end{equation}

The range of output solution distributions associated with these mean values are shown in \ref{eq17}:

\begin{equation} \label{eq17}
[x]_{Range} = \Bigg[ \begin{smallmatrix} 0 \leq \ x_{MS} \  \leq 2.5\times10^{-4} \\ 0\leq \ x_{SA} \  \leq \ 2.5\times10^{-4} \\ 0\leq \  x_{MA} \ \leq \  4.5\times10^{-3} \\x_{Si}\ = \  1.2\times10^{-7} \pm \  3 \times10^{-8}\end{smallmatrix} \Bigg]
\end{equation}
The solution values for $x_{MS}$, $x_{SA}$, and $x_{MA}$ are approximately evenly distributed throughout the ranges shown in \ref{eq17}.  $x_{Si}$ exhibits an approximately Guassian distribution with a standard deviation given by the uncertainty given in \ref{eq17}. \par

\begin{table*}
\renewcommand\arraystretch{2}
\begin{tabular}{|| c  c  c  c  c  c  c ||} 
\hline
$\tan{\delta}$ & This work & Ref. \onlinecite{sOConnell2008} & Ref. \onlinecite{sQuintana2014} & Ref. \onlinecite{sGambetta2017} & Ref. \onlinecite{sWenner2011} & Ref. \onlinecite{sWang2015}\\
\hline \hline
$\tan{\delta_{MS}} $& $5.9\times10^{-4}$ & - & - & - & - & $ <2.6\times10^{-3}$\\
\hline
$\tan{\delta_{SA}} $& $7.1\times10^{-4}$& $3.1\times10^{-4}$ & - & - & - & $ <2.2\times10^{-3}$\\
\hline
$\tan{\delta_{MA}}$ & $3.9\times10^{-3}$& $1.5\times10^{-3}$ & $2.6\times10^{-3}$ & - & $2\times10^{-3}$ & $2.1\times10^{-2}$\\
\hline
$\tan{\delta_{Si}}$ & $1.2\times10^{-7}$& - & - & $<5\times10^{-7}$& - & $<1\times10^{-6}$\\
\hline
\end{tabular}
\caption{Comparisons between reported interface dielectric loss tangents.}
\label{tab}
\end{table*}

The uncertainty in these values results from multiple contributions.  First, the error in the estimation of the mean $Q_{TLS}$ values derived from measurement statistics results in a range of possible solutions to the set of linear equations Eq. \ref{eq6}.  Second, the nearly singular nature of the matrix [P] prevents a unique solution from being identified.  Finally, although we have sought to minimize these contributions, systematic errors such as incomplete characterization of the device geometry or residual contributions of $Q_{HP}$ to $Q_{TLS}$  may contribute to uncertainty in the loss factor solution.  Nevertheless, the correlation between the measured and predicted $Q_{TLS}$  shown in Fig. 2(c) of the main text in the main text demonstrates that these values have predictive power within a tolerance set by the loss factor uncertainty.\par

In order to convert these values to true loss tangents, the extracted loss factors can be combined with Eqs. \ref{eq3} and \ref{eq4} using the nominal dielectric interface parameters used in simulations ($\epsilon_{nom}$ = 10$\epsilon_0$ and t$_{nom}$ = 10 nm) and reasonable assumptions for the properties of each interface: th$_{MS}$ = 2 nm, t$_{SA}$ = 2 nm, t$_{MA}$ = 2 nm, $\epsilon_{MS}$ = 11.7$\epsilon_0$, $\epsilon_{SA}$ = 4$\epsilon_0$, $\epsilon_{MA}$=10$\epsilon_0$.  The resulting mean best-fit loss tangents are given in Eq. \ref{eq18}:
\begin{equation} \label{eq18}
\bigg[ \begin{smallmatrix} \tan{\delta_{MS}} \\ \tan{\delta_{SA}} \\ \tan{\delta_{MA}}  \\ \tan{\delta_{Si}}\end{smallmatrix} \bigg]_{Mean} = \Bigg[ \begin{smallmatrix} 5.9\times10^{-4} \\7.1\times10^{-4} \\3.9\times10^{-3} \\1.2\times10^{-7}\end{smallmatrix} \Bigg]
\end{equation}
The ranges of output corresponding to these mean values are shown in \ref{eq19}:
\begin{equation} \label{eq19}
[\tan{\delta}]_{Range} = \Bigg[ \begin{smallmatrix} 0 \leq \ \tan{\delta_{MS}} \  \leq 1.5\times10^{-3} \\ 0\leq \ \tan{\delta_{SA}} \  \leq \ 3.1\times10^{-3} \\ 0\leq \  \tan{\delta_{MA}} \ \leq \  2.3\times10^{-2} \\ 1.2\times10^{-7} \pm \  3 \times10^{-8}\end{smallmatrix} \Bigg]
\end{equation}
Despite the uncertainty in the loss factors, dielectric interface thicknesses and dielectic constants, the best-fit loss tangent values all fall within the reported range of loss tangents for these materials and interfaces from the literature, as shown in Table \ref{tab}.\par

\end{document}